\documentclass[aps,prb,groupedaddress,twocolumn,floatfix]{revtex4}

\newcommand{\lambdab}{{\mbox{\boldmath{$\lambda$}}}} 
\newcommand{\Gammab}{{\mbox{\boldmath{$\Gamma$}}}}
 
\usepackage{graphicx}
\usepackage[]{amsmath}

\begin{document}


\title{An analytical treatment of in-plane magnetotransport in the Falicov-Sievert model}


\author{Andrzej Nowojewski} \affiliation{Department of Physics,
  Harvard University, Cambridge, Massachusetts 02138, USA}
\author{Stephen J. Blundell} \altaffiliation{Corresponding author:
  s.blundell@physics.ox.ac.uk} \affiliation{Clarendon Laboratory,
  Department of Physics, University of Oxford, Parks Road, Oxford OX1
  3PU, United Kingdom}

\date{\today}

\newcommand{\chem}[1]{\ensuremath{\mathrm{#1}}}

\begin{abstract}
We derive an analytical expression which allows efficient computation
of the effect of all the Fermi surface trajectories induced by a
combination of Bragg scattering and magnetic breakdown on the in-plane
components of the resistivity tensor.  The particular network of
coupled orbits which we consider was first formulated by Falicov and
Sievert, who studied the problem numerically.  Our approach, based
upon a method used previously to derive an analytical solution for
interlayer transport, allows us to show that the conductivity tensor
can be written as a sum of a matrix representing the effect of total
magnetic breakdown and one representing a combination of complex
electronic trajectories, and we find a compact expression for the
in-plane components of the resistivity tensor that can be evaluated
straightforwardly.
\end{abstract}

\maketitle


\section{Introduction}
Magnetoresistance has long been used as a tool to study the Fermi
surface of metals \cite{Shoenberg1984,Pippardbook}.  The observation
of quantum oscillations provides a measure of the cross-sectional area
of closed pockets (due to the quantization of the Landau levels).  The
background magnetoresistance also contains information about all the
electronic orbits induced by the magnetic field.  For example, a
transverse magnetoresistance saturating in high field can be evidence
for the presence of closed orbits, while a magnetoresistance that
increases quadratically with field can indicate open orbits.  A number
of effects can complicate this picture and one such is magnetic
breakdown, a concept first introduced to explain a giant orbit
observed in magnesium \cite{cohenfalicov}.  Because of the periodic
potential in a crystal, small gaps in the dispersion sometimes open up
at the Brillouin zone edge, splitting the Fermi surface into distinct
sections.  In low magnetic fields, electrons traverse the Fermi
surface under the action of the magnetic field and Bragg scatter at
the Brillouin zone edge.  In a sufficiently strong magnetic field it
is possible for an electron to tunnel from one segment of the Fermi
surface to another \cite{cohenfalicov,blount}.  This is because in
high magnetic field, mixing of states between different pieces of the
Fermi surface can increase the likelihood of magnetic breakdown, so
that the resulting orbit more closely resembles that which would be
obtained in the absence of the periodic potential \cite{pippard62}.

The presence of magnetic breakdown presents a challenge for the
simulations of magnetoresistance.  This is because at the points on
the Brillouin zone edge intersected by the Fermi surface (points which
we call magnetic breakdown junctions) there is a probability $p$ for
magnetic breakdown to occur and this is given by $ p=\exp(-B_0/B)$,
where $B_0$ is the characteristic breakdown field; the problem is that
under usual experimental conditions $p$ is {\it between} 0 and 1.
Thus there is a hierarchy of complex trajectories that must be summed
to account for all possible contributions to the conductivity in which
magnetic breakdown either does or does not occur at each magnetic
breakdown junction.  If a quasiparticle crosses $N$ magnetic breakdown
junctions, one has to consider $2^N$ possible trajectories with their
correct probabilistic weightings, and this complicates a direct
computation of magnetoresistance since one has to sum over
trajectories with arbitrarily long path lengths and hence arbitarily
large values of $N$.

In order to calculate how magnetic breakdown can change the
connectivity of the orbits \cite{falicovsievert64}, and thus the
resistivity, Falicov and Sievert in 1965 provided a model
\cite{falicovsievert} that generalised the Chambers path integral
method \cite{chambers} to include magnetic breakdown at a finite
number of points.  This was used to evaluate the magnetoresistance in
a number of model Fermi surfaces, although the calculations were
performed numerically.

More recently, the problem of the effect of magnetic breakdown on the
magnetoresistance in a network of coupled orbits was treated
analytically \cite{andrzej}, though the focus of this work was on
interlayer transport.  This is because the newer calculations are
motivated by recent experiments on quasi-two-dimensional organic
metals \cite{Goddard2004,kang,bamroexp,kang09} in which
magnetotransport experiments have been particularly illuminating
\cite{Kartsovnik2004}.  In this paper we use the techniques developed
in Ref.~\onlinecite{andrzej} (see also Ref.~\onlinecite{niseko}) to
derive an analytical expression for the in-plane components of the 
resistivity tensor in the
Falicov-Sievert model.  This expression allows the underlying physics
of the magnetotransport to be more clearly extracted and will allow a
more convenient comparison with experimental data.

\begin{figure}
\includegraphics[width=9.0cm]{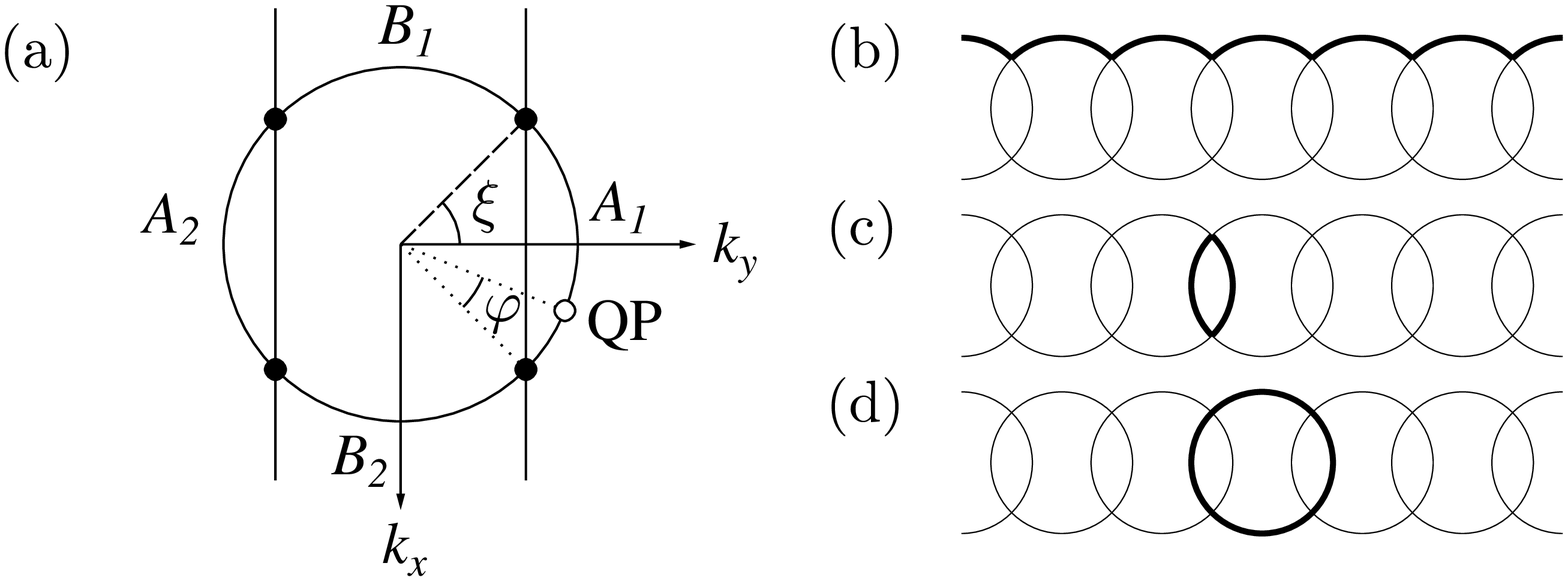}
\caption{(a) The Fermi surface in the $k_x$-$k_y$ plane showing the
  points where magnetic breakdown can occur which are at 
$(k_x,k_y)=(\pm k_{\rm F}\sin\xi,\pm k_{\rm F}\cos\xi)$ (these are
  called
magnetic breakdown junctions).  The
azimuthal coordinate of a quasiparticle at the point labelled QP is 
$\varphi$.
(b) An open orbit (comprising the repeated traversal of the $B_1$
  section).  
(c) Closed orbit (the $\alpha$-orbit,
comprising the repeated traversal of $A_1$ and 
$A_2$).
(d) Breakdown orbit (the $\beta$-orbit, comprising
  $A_1$-$B_1$-$A_2$-$B_2$).  
}
\label{fig:fs}
\end{figure}

\section{Introduction to the model}
In this paper, we consider the quasi-two-dimensional Fermi surface
shown in Fig.~\ref{fig:fs}(a) in which the dispersion is given by
$E(k)={ \hbar^2(k_x^2+k_y^2) \over 2m^*}-2t_\perp\cos k_zd_\perp$,
where $m^*$ is the effective mass, $k_{\rm F}$ is the Fermi wave
vector, $d_\perp$ is the interlayer spacing and the interlayer hopping
$t_\perp$ is small ($t_\perp\ll \hbar k_{\rm F}/d_\perp$).  The Fermi
surface consists of a cylinder with volume $\pi k_{\rm
  F}^2 \times \frac{2\pi}{d_{\perp}}$ and thus the number density of
electrons is $n=k_{\rm F}^2/2\pi d_\perp$.  The Brillouin zone edges
are placed at $k_y=\pm k_{\rm F}\cos\xi$.  Quasiparticles orbit around
the Fermi surface with constant $k_z$ when $B$ lies along the
interlayer direction. In very low $B$, because of Bragg reflection,
only open orbits [Fig.~\ref{fig:fs}(b)] and small closed orbits [the
  $\alpha$ orbit, Fig.~\ref{fig:fs}(c)] occur around the distinct
sections of the Fermi surface.  In high $B$, mixing between the states
on the two Fermi surface sections leads to magnetic breakdown at the
four magnetic breakdown junctions.  At these junctions a quasiparticle
``tunnels'' in $k$-space between the Fermi surface sections, resulting
in a single large closed orbit [the $\beta$ orbit,
  Fig.~\ref{fig:fs}(d)].  This model is equivalent to the situation
outlined in Fig.~1 of Ref.~\onlinecite{falicovsievert} but the
notation has been chosen to link with the physical situation relevant
to a particular family of organic metals \cite{Goddard2004,andrzej}.

\section{Magnetoresistance with no Bragg scattering}
In this section we review the standard calculation of
magnetoresistance in the case of full magnetic breakdown (i.e.\ $p=1$) which is
equivalent to assuming a very high magnetic field and ignoring any
Bragg scattering.  In this case we consider only the effect of a
single closed breakdown orbit [i.e.\ the case of Fig.~\ref{fig:fs}(d)].
The Boltzmann equation gives
the conductivity tensor
\begin{equation}
\sigma_{\alpha\beta} = e^2\tau g(E_{\rm F}) \langle v_\alpha
\bar{v_\beta} \rangle_{\rm FS}.
\end{equation}
as an average of velocity correlations over the Fermi surface, where
$\bar{v_\beta}$ is given by 
\begin{equation}
\bar{v_\beta}=\int_0^\infty
{{\rm e}^{-t/\tau}\over \tau}v_\beta[{\bf k}(t)]\,{\rm d}t,
\end{equation}
and 
$g(E) = {m^* / d_\perp \pi \hbar^2}$ is the density of states.
The quasiparticle orbits lie in a plane perpendicular to 
the magnetic field ${\bf B}$ and orbit the Fermi surface with 
angular frequency $\omega_{\rm c}$ given by 
$\omega_{\rm
  c}={ eB/ m^*}$ if ${\bf B}$ is perpendicular to the
quasi-two-dimensional planes.
Writing $\varphi=\omega_{\rm c} t + \varphi_0$, we have that
\begin{equation}
\bar{v_\beta} = {1 \over\omega_{\rm c}\tau}
\int_{\varphi_0}^\infty v_\beta(k) {\rm
  e}^{-(\varphi-\varphi_0)/\omega_{\rm c}\tau}\,{\rm d}\varphi
\end{equation}
and so
the conductivity tensor can be written as
\begin{equation}
\sigma_{\alpha\beta} = {\sigma_0 \over v_{\rm F}^2 \pi \gamma}
\int_0^{2\pi} {\rm d}\varphi_0\, v_\alpha
\int_{\varphi_0}^\infty v_\beta(k) {\rm
  e}^{-(\varphi-\varphi_0)/\gamma}\,{\rm d}\varphi
\label{eq:sigma1}
\end{equation}
where 
\begin{equation}
\gamma = \omega_{\rm c}\tau = {e B \tau \over m^*}
\end{equation}
and 
\begin{equation}
\sigma_0 = {e^2 m^* v_{\rm
  F}^2 \tau \over  2 \pi \hbar^2 d_\perp} = {n e \gamma \over  B}.
\end{equation} 
Neglecting the small effect of the 
interplane warping $t_\perp$ on orbit size, and considering only the
in-plane components of the conductivity tensor
$\mbox{\boldmath{$\sigma$}}$, we proceed as follows.
Setting $v_x=v_{\rm F}\cos\varphi$ and $v_y=v_{\rm F}\sin\varphi$, and
assuming full magnetic breakdown (so all the orbits just go round the
cylinder with no Bragg scattering), the integral can be evaluated to
give
\begin{equation}
\mbox{\boldmath{$\sigma$}} 
= 
{\sigma_0 \over 1+\gamma^2} \left( \begin{matrix} 1 & \gamma \cr
  -\gamma & 1 \end{matrix} \right)
\label{eq:simplesigma}
\end{equation}
and inverting this tensor yields the resistivity tensor 
\begin{equation}
\mbox{\boldmath{$\rho$}}  = \mbox{\boldmath{$\sigma$}}^{-1} = \sigma_0^{-1}
 \left( \begin{matrix} 1 & -\gamma \cr \gamma & 1 \end{matrix}
 \right),
\label{eq:simplerho}
\end{equation}
or equivalently
\begin{equation}
\mbox{\boldmath{$\rho$}}  = 
 \left( \begin{matrix} \sigma_0^{-1} & -B/ne \cr B/ne & \sigma_0^{-1} \end{matrix} \right)
\end{equation}
This result is exactly what one expects from the simple theory of metals:
no magnetoresistance (because $\rho_{xx}=\rho_{yy}= \sigma_0^{-1}$ is
not field-dependent) and a simple Hall effect $\rho_{xy}=-\rho_{yx} =
R_{\rm H} B$ where $R_{\rm H}=1/ne$.
For later reference, we will denote the simple form of the resistivity
tensor with no magnetic breakdown in Eq.~(\ref{eq:simplerho}) as
$\mbox{\boldmath{$\rho$}}^0$.

\section{Magnetoresistance including magnetic breakdown and Bragg scattering}
A treatment fully including both magnetic breakdown and Bragg
scattering can be performed using
the method described in Ref.~\onlinecite{andrzej}.
Setting $\nu_{\alpha}=v_{\alpha}/v_F$ with $\nu_x=\cos\varphi$ and
$\nu_y=\sin\varphi$,
Eq.~(\ref{eq:sigma1}) becomes
\begin{equation}
\sigma_{\alpha\beta} = {\sigma_0\over\pi\gamma}
\int_0^{2\pi} {\rm d}\varphi_0\, \nu_\alpha {\rm
  e}^{\varphi_0/\gamma}
\int_{\varphi_0}^\infty \nu_\beta(k) {\rm
  e}^{-\varphi/\gamma}\,{\rm d}\varphi.
\end{equation}
and with the definitions of the functions ${\rm E}_{\alpha}^{\pm}(x)$
\begin{equation}
{\rm E}_{\alpha}^{\pm}(x)=\nu_\alpha(x) {\rm
  e}^{\pm x/\gamma},
\end{equation}
the conductivity can be written
\begin{equation}
\sigma_{\alpha\beta} = {\sigma_0\over\pi\gamma}
\int_0^{2\pi} {\rm d}\varphi_0\, {\rm E}_{\alpha}^{+}(\varphi_0)
\int_{\varphi_0}^\infty {\rm E}_{\beta}^{-}(\varphi)\,{\rm
  d}\varphi.
\label{eformula}
\end{equation}
 What makes
Eq.~(\ref{eformula}) challenging to evaluate is that the integrand
changes depending on the path taken by the quasiparticle which, at
each magnetic breakdown junction of the orbit, can either undergo magnetic breakdown tunneling (with
probability $p\equiv {\rm e}^{-B_0/B}$) or Bragg reflection
(with probability $q=1-p$)\cite{pippard62,Shoenberg1984}. 
The strategy for solving this problem follows the earlier approach
\cite{andrzej,niseko} of 
separating the motion of electrons in the four different segments
of the orbit and constructing recursive equations which encode all the
information about the behaviour at the magnetic breakdown junctions.  
In this way we can write the conductivity tensor as
\begin{eqnarray}
\sigma_{\alpha\beta}= {\sigma_0\over\pi\gamma}\lambdab_{\alpha}^{+}\cdot\left(\lambdab_{\beta}^{\rm init}+
\Gammab(\boldsymbol{I}-\Gammab
)^{-1}\cdot\lambdab_{\beta}^{-}\right),
\label{eq:mainsigma}
\end{eqnarray}
where the vectors $\lambdab^{\pm}_\alpha$ and $\lambdab^{\rm
  init}_\alpha$ are given by
\begin{eqnarray}
\lambdab_{\alpha}^{\pm}=\left(\begin{array}{c}\int_{0}^{2\xi}dx
E_{\alpha}^{\pm}(x)\\a^{\pm1}\int_{2\xi}^{\pi}dx
E_{\alpha}^{\pm}(x)\\(ab)^{\pm1}\int_{\pi}^{\pi+2\xi}dx
E_{\alpha}^{\pm}(x)\\(a^{2}b)^{\pm1}\int_{\pi+2\xi}^{2\pi}dx
E_{\alpha}^{\pm}(x)\end{array}\right),
\label{eq:lambdaalpha}
\end{eqnarray}
and
\begin{eqnarray}
\lambdab_{\alpha}^{\textrm{init}}=\left(\begin{array}{c}\int_{\varphi_0}^{2\xi}d\varphi
  E_{\alpha}^{-}(\varphi)\\a^{-1}\int_{\varphi_0}^{\pi}d\varphi
  E_{\alpha}^{-}(\varphi)\\(ab)^{-1}\int_{\varphi_0}^{\pi+2\xi}d\varphi
  E_{\alpha}^{-}(\varphi)\\(a^{2}b)^{-1}\int_{\varphi_0}^{2\pi}d\varphi
  E_{\alpha}^{-}(\varphi)\end{array}\right),
\label{eq:lambdainit}
\end{eqnarray}
and the matrix $\Gammab$ which encodes all the breakdown probabilities
and scattering is given by
\begin{equation}
\Gammab=\left(
\begin{array}{cccc}
a & 0 & 0 & 0 \\
0 & b & 0 & 0 \\
0 & 0 & a & 0 \\
0 & 0 & 0 & b \\
\end{array} \right)\cdot\left(
\begin{array}{cccc}
0 & p & q & 0 \\
0 & q  & p & 0 \\
q & 0 & 0 & p \\
p & 0 & 0 & q  \\
\end{array} \right),
\end{equation}
where
$a=\exp\left(-{2\xi}/{\gamma}\right)$ and
$b=\exp\left(-{(\pi-2\xi)}/{\gamma}\right)$.

\begin{figure}[htb]
\centering
\includegraphics[width=9.0cm]{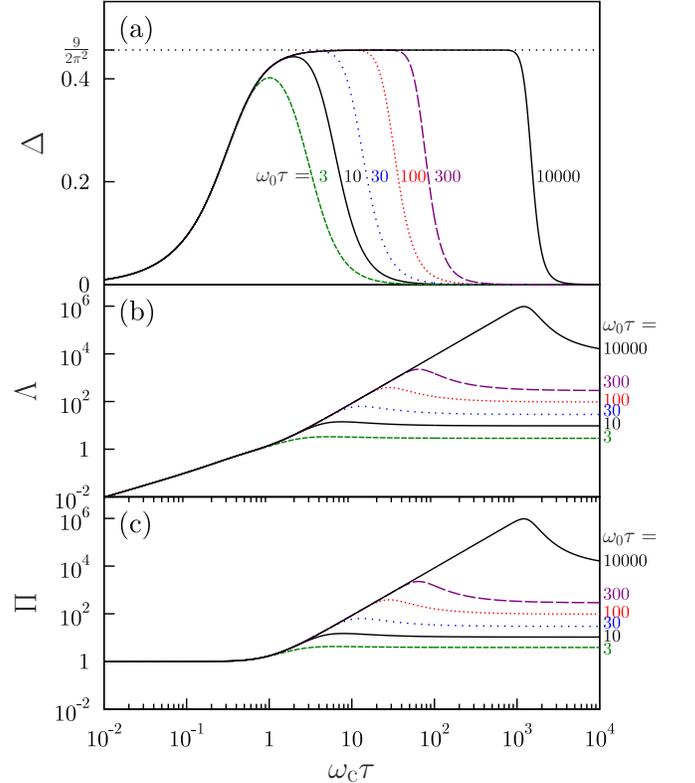}
\caption{
(Color online.)
The parameters (a) $\Delta$ [Eq.~(\ref{eq:delta})], (b) $\Lambda$ [Eq.~(\ref{eq:2})]
and (c) $\Pi$ [Eq.~(\ref{eq:3})].
}
\label{fig:delta}
\end{figure}

The integrals can be performed analytically and after much
simplification (see Appendix) the conductivity can be obtained as
\begin{equation}
\mbox{\boldmath{$\sigma$}} 
={\sigma_0\over1+\gamma^2}\left(
\begin{array}{cc}
 1 & \gamma     \\
 -\gamma &  1\end{array}
\right)+ 
\Delta
{2\sigma_0\over 1+\gamma^2}
{\bf X},
\label{conduc}
\end{equation}
where the parameter $\Delta$ is given by
\begin{eqnarray}
\Delta={4\vert\gamma\vert\over\pi(1+\gamma^2)}
\left(
{q(1+ab)\cos^2\xi\over 1-bq+a(q+b(p^2-q^2))}
\right),
\label{eq:delta}
\end{eqnarray}
and the matrix ${\bf X}$ is given by
\begin{equation} 
{\bf X}
=
\left(
\begin{array}{cc}
 \gamma^2\sin^2\xi-\cos^2\xi & 
- \gamma - (\gamma^2+1)\sin\xi\cos\xi \\
 \gamma - (\gamma^2+1)\sin\xi\cos\xi &
 \gamma^2\cos^2\xi-\sin^2\xi\end{array}
\right).
\end{equation}
Equation~(\ref{conduc}) demonstrates that the conductivity is a sum of the
familiar expression assuming no Bragg scattering 
[i.e.\ Eq.~(\ref{eq:simplesigma})]
and an additional
term.  This additional term is included by an amount controlled by
$\Delta$, and it is therefore useful to plot the field dependence of
this quantity.  This is shown in Fig.~\ref{fig:delta}(a)
for the case in which $\xi=\pi/6$ (chosen to match Fig.~1 of
Ref.~\onlinecite{falicovsievert}) for a range of different breakdown
fields
(parametrized by $\gamma_0 \equiv \omega_0 \tau = eB_0\tau/m^*$)
and plotted as a function of magnetic field
(parametrized by $\gamma \equiv \omega_{\rm c} \tau = eB\tau/m^*$).

For low field $\Delta \approx \alpha \gamma/(1+\gamma^2)$
where $\alpha={(4\cos^2\xi)/\pi}$ and is independent of breakdown field
$\gamma_0$. For intermediate fields, and for $\gamma_0 \gg 1$, $\Delta$
rises to a plateau given by
$\Delta \approx 4\cos^2\xi / [\pi (\pi-2\xi)] = \alpha/ (\pi-2\xi)$.
In our example in which $\xi=\pi/6$, the plateau is
at $\Delta \approx 9/2\pi^2$ [see Fig.~\ref{fig:delta}(a)]. 
For large fields, $\Delta$ decreases again and 
follows $\Delta \approx \alpha  \gamma_0 / \gamma^2$.
Thus we expect that at very low and very high fields the approximation
of no Bragg scattering (the single breakdown orbit) in
Eq.~(\ref{eq:simplesigma}) will work well.  It works at very low
fields because electrons travel only a very short distance around the
Fermi surface before scattering and so orbit connectivity is largely irrelevant.
It works at very high fields because then the magnetic breakdown probability is close
to unity.  At intermediate fields $\Delta$ is significant and the
effect of magnetic breakdown and Bragg scattering is important in determining the magnetotransport.

\begin{figure}[htb]
\centering
\includegraphics[width=9.0cm]{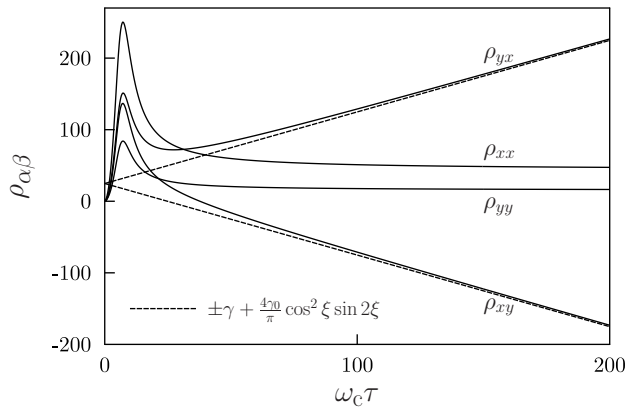}
\caption{
The components of the resistivity tensor (plotted in units of $\sigma_0^{-1}$)
calculated according to
Eqs.~(\ref{eq:1}--\ref{eq:3})
for the case of $\xi=\pi/6$ and $\omega_0\tau=30$.
}
\label{fig:basic}
\end{figure}

\begin{figure*}[htb]
\centering
\includegraphics[width=\textwidth]{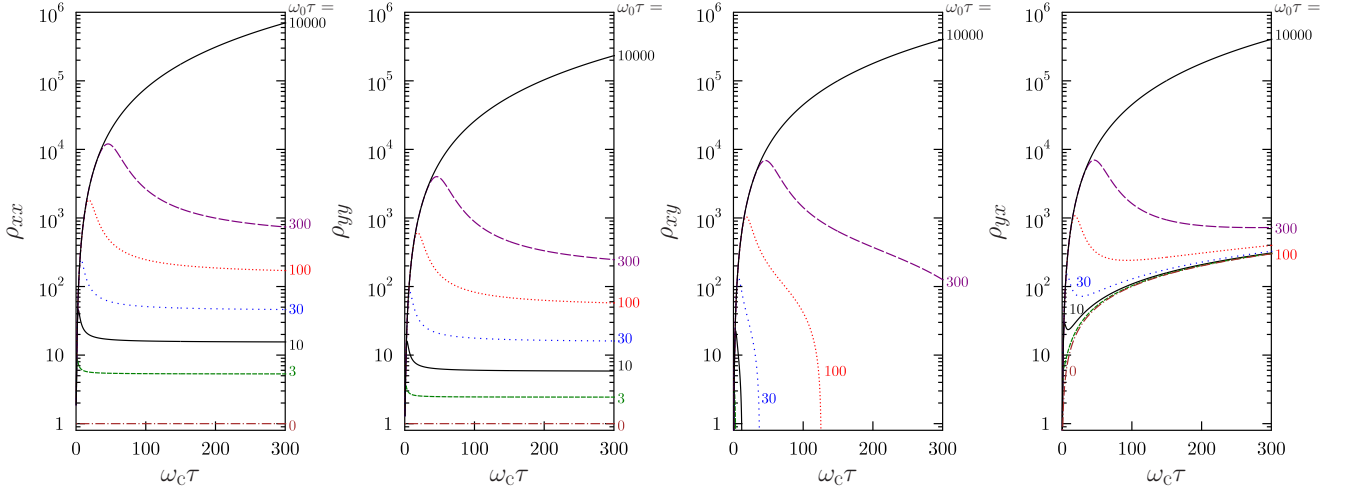}
\caption{
(Color online.)
The components of the resistivity tensor  (plotted in units of
$\sigma_0^{-1}$)
calculated according to
equations~(\ref{eq:1}--\ref{eq:3})
for the case of $\xi=\pi/6$.
}
\label{fig:rho}
\end{figure*}

We now invert the conductivity tensor in
Eq.~(\ref{conduc}) to obtain an expression for the resistivity
tensor, and find
\begin{equation}
\mbox{\boldmath{$\rho$}} = \sigma_0^{-1} 
\left(
\begin{array}{cc}
\Pi + \Lambda\cos 2\xi & -\gamma + \Lambda \sin 2\xi \\
\gamma + \Lambda \sin 2\xi & \Pi - \Lambda \cos 2\xi
\end{array}
\right),
\label{eq:1}
\end{equation}
where
\begin{equation}
\Lambda = { \Delta (\gamma^2 + 1) \over 1 - 2\Delta }
\label{eq:2}
\end{equation}
and 
\begin{equation}
\Pi = { 1 + \Delta (\gamma^2 - 1) \over 1- 2\Delta }.
\label{eq:3}
\end{equation}
Equations~(\ref{eq:1}--\ref{eq:3}) are the main results of the paper.
The quantities $\Lambda$ and $\Pi$ are plotted in
Fig.~\ref{fig:delta}(b) and (c).
In low field ($\gamma\ll 1$) $\Lambda \approx \alpha \gamma$ and 
$\Pi \approx 1$.  In large field ($\gamma\gg 1$) $\Lambda \approx
\alpha \gamma_0$ and $\Pi \approx 1 + \alpha\gamma_0$. 
These results can be used to show that in low field the resistivity
tensor
is
\begin{equation}
\mbox{\boldmath{$\rho$}}
=
\mbox{\boldmath{$\rho$}}^0
+
\sigma_0^{-1}\gamma\alpha
{\bf Y},
\end{equation}
while in high
fields the resistivity tensor is
\begin{equation}
\mbox{\boldmath{$\rho$}}
=
\mbox{\boldmath{$\rho$}}^0
+
\sigma_0^{-1}\gamma_0\alpha
{\bf Y},
\label{highfieldlimit}
\end{equation}
where the matrix ${\bf Y}$ contains only geometric
factors and is given by
\begin{equation}
{\bf Y}=
\left(
\begin{array}{cc}
 1+\cos2\xi & \sin2\xi    \\
 \sin2\xi &   1-\cos2\xi
\end{array}
\right).
\end{equation}

\section{Discussion}
The components of the resistivity tensor obtained using
equations~(\ref{eq:1}--\ref{eq:3}) are shown in Fig.~\ref{fig:basic}
for the case of $\xi=\pi/6$ and $\omega_0\tau=30$.  At high magnetic
field the diagonal components of the resistivity tensor saturate (to a
value linear in $\gamma_0$), while the off-diagonal terms approach the
asymptotic values proportional to $\pm\gamma+\gamma_0\alpha\sin 2\xi$
[in agreement with Eq.~(\ref{highfieldlimit})].  As the magnetic field
is decreased all resistivity tensor elements go through a maximum that
becomes sharper with increasing breakdown field $\gamma_0$.  At lower
fields they decrease linearly.  The position of the maximum is
controlled predominantly by the behavior of $\Delta$ plotted in
Fig.~\ref{fig:delta}.

The components of the resistivity tensor are plotted again in
Fig.~\ref{fig:rho}, though this time for a range of values of
breakdown fields (parametrized by $\gamma_0=\omega_0\tau$). [The plot
  of $\rho_{xx}$ is identical to that of Fig.~1(c) in
  Ref.~\onlinecite{falicovsievert}, demonstrating the agreement of the
  analytical expressions in Eqs.~(\ref{eq:1}--\ref{eq:3}) and the
  earlier numerical work.]  The difference between $\rho_{xx}$ and
$\rho_{yy}$ reflects the fact that for our considered geometry in
Fig.~\ref{fig:rho} ($\xi=\pi/6$) the open orbits which occur due to
Bragg scattering are efficient at carrying current in the
$y$-direction.  This difference vanishes when $\xi=\pi/4$.

It is worthwhile to show that the off-diagonal components of the resistivity
tensor do indeed obey Onsager symmetry,
i.e.\ $\rho_{xy}(B)=\rho_{yx}(-B)$.  Onsager
symmetry is dependent upon perfect microscopic reversibility (with the
appropriate sign change in the magnetic field)\cite{onsager}, and we
find that this
is obeyed even
with the presence of magnetic breakdown junctions, despite the apparent
``randomization'' of the electron trajectory which occurs at magnetic
breakdown junctions. 
Reversal of the magnetic field changes the way in which the sections
of the Fermi surface are connected at the magnetic breakdown junctions
and the result of this is that the connectivity matrix $\Gamma$
transforms into its transpose.  This has the effect of making $\Delta$
invariant under a sign change of the magnetic field [and is the origin
of the modulus sign of the factor of $\gamma$ in the numerator of
Eq.~(\ref{eq:delta})]. The net consequence of this is that 
Onsager symmetry is preserved.

In summary, in this paper we have derived an analytical expression for
the in-plane components of the resistivity tensor that includes the
effect of all the orbits induced by a combination of Bragg scattering
and magnetic breakdown.  Eq.~(\ref{conduc}) demonstrates that approach
shows that the conductivity tensor can be written as a sum of a matrix
representing the effect of total magnetic breakdown and one
representing a combination of complex electronic trajectories.  Our
main result in Eqs.~(\ref{eq:1}--\ref{eq:3}) provides a compact form
for the in-plane components of the resistivity tensor for this problem
which are in a convenient form to compare with experimental data.

\section*{Acknowledgments}
We are grateful to EPSRC (UK) and Mansfield College, Oxford
for financial support.

\section*{Appendix}
The expression for conductivity from Eq.~(\ref{eq:mainsigma}) is 
\begin{eqnarray}
\sigma_{\alpha\beta}=
      {\sigma_0\over\pi\gamma}\lambdab_{\alpha}^{+}\cdot\left(\lambdab_{\beta}^{\rm
        init}+
\Gammab(\boldsymbol{I}-\Gammab
)^{-1}\cdot\lambdab_{\beta}^{-}\right),
\end{eqnarray}
and as in Ref.~\onlinecite{andrzej} the matrix
$\Gammab(\boldsymbol{I}-\Gammab )^{-1}$ is given by
\begin{eqnarray}
\Gammab(\boldsymbol{I}-\Gammab )^{-1}=\frac{1}{N}\left( \begin{array}{cccc}
t & ap r & ars & a^{2}p s \\
abp s & w & bp r & abp^{2} \\
a rs & a^{2}p s & t & ap r \\
bp r & abp^{2} & abp s & w \\
\end{array} \right),\nonumber
\end{eqnarray}
where
$r=1-bq$, $s=q+b(p^{2}-q^{2})$, $N=r^2-a^2s^2$, $t=a^2s^2$ and $w=b(qr+a^{2}(p^{2}-q^{2})s)$.
The function $E_\alpha^\pm(x)$ can be simplified because
$\nu_x=\cos\varphi$, $\nu_y=\sin\varphi$, so that
\begin{eqnarray}
{\rm E}_{x}^{\pm}(\varphi)=\cos\varphi {\rm
  e}^{\pm \varphi/\gamma}=\Re\left({\rm
  e}^{(\imath\pm 1/\gamma)\varphi}\right)\nonumber\\
  {\rm E}_{y}^{\pm}(\varphi)=\sin\varphi {\rm
  e}^{\pm \varphi/\gamma}=\Im\left({\rm
  e}^{(\imath\pm 1/\gamma)\varphi}\right).
\end{eqnarray}
The vector in Eq.(\ref{eq:lambdaalpha}) can then be written
\begin{equation}
\lambdab_{\alpha}^{\pm}=\left(\begin{array}{c}\lambda^1_{\alpha\pm}\\\lambda^2_{\alpha\pm}\\\lambda^3_{\alpha\pm}\\\lambda^4_{\alpha\pm}\end{array}\right)
=\left(\begin{array}{c}z^{\pm}\left({\rm e}^{2\imath\xi}  a^{\mp
    1}-1\right)\\-z^{\pm}\left({\rm e}^{2\imath\xi} +b^{\mp
    1}\right)\\-\lambda^1_{\alpha\pm}\\-\lambda^2_{\alpha\pm}\end{array}\right),
\label{lambdaalphapm}
\end{equation}
and that in Eq.(\ref{eq:lambdainit}) as
\begin{equation}
\lambdab_{\alpha}^{\rm init}=\left(\begin{array}{c}\lambda^1_{\alpha
    i}\\\lambda^2_{\alpha i}\\\lambda^3_{\alpha i}\\\lambda^4_{\alpha
    i}\end{array}\right)
=\left(\begin{array}{c}z^{-} {\rm e}^{2\imath\xi}
  a\\-z^{-}b\\-\lambda^1_{\alpha i}\\-\lambda^2_{\alpha
    i}\end{array}\right),
\label{lambdaalphainit}
\end{equation}
with $z^{\pm}=\left(\imath\pm1/\gamma\right)^{-1}$.
The two first and two last entries of each vector
differ by a minus sign because these parts of the orbit are exactly
$\pi$ apart. With this identity 
we can readily calculate:
\begin{widetext}
\begin{equation}
{1\over 2}\lambda_{\alpha}^{+} \cdot\Gammab(\boldsymbol{I}-\Gammab
)^{-1}\cdot\lambda_{\beta}^{-} =
-(\lambda^{1}_{\alpha+}\lambda^{1}_{\beta-}+\lambda^{2}_{\alpha+}\lambda^{2}_{\beta-})+\frac{\lambda^{1}_{\alpha+}\lambda^{1}_{\beta-}(1-bq)+\lambda^{2}_{\alpha+}\lambda^{2}_{\beta-}(1+aq)+\lambda^{1}_{\alpha+}\lambda^{2}_{\beta-}ap-\lambda^{2}_{\alpha+}\lambda^{1}_{\beta-}bp}{1-bq+a(q+b(p^2-q^2))}
\end{equation}
The initial contribution can be simplified as follows:
\begin{align}
\lambda_{\alpha}^{+}\cdot\lambda_{\beta}^{\textrm{init}}&=\left(\begin{array}{c}\lambda^1_{\alpha+}\\\lambda^2_{\alpha+}\\\lambda^3_{\alpha+}\\\lambda^4_{\alpha+}\end{array}\right)\cdot\left(\begin{array}{c}\lambda^1_{\beta i}\\\lambda^2_{\beta i}\\\lambda^3_{\beta i}\\\lambda^4_{\beta i}\end{array}\right)\nonumber\\&=\left(\begin{array}{c}\int_{0}^{2\xi}d\varphi_0
E_{\alpha}^{+}(\varphi_0)\\a\int_{2\xi}^{\pi}d\varphi_0
E_{\alpha}^{+}(\varphi_0)\\ab\int_{\pi}^{\pi+2\xi}d\varphi_0
E_{\alpha}^{+}(\varphi_0)\\a^{2}b\int_{\pi+2\xi}^{2\pi}d\varphi_0
E_{\alpha}^{+}(\varphi_0)\end{array}\right)\cdot\left(\begin{array}{c}\int_{\varphi_0}^{2\xi}d\varphi
E_{\beta}^{-}(\varphi)\\a^{-1}\int_{\varphi_0}^{\pi}d\varphi
E_{\beta}^{-}(\varphi)\\(ab)^{-1}\int_{\varphi_0}^{\pi+2\xi}d\varphi
E_{\beta}^{-}(\varphi)\\(a^{2}b)^{-1}\int_{\varphi_0}^{2\pi}d\varphi
E_{\beta}^{-}(\varphi)\end{array}\right)\nonumber\\&=\left(\begin{array}{c}\int_{0}^{2\xi}d\varphi_0
E_{\alpha}^{+}(\varphi_0)\\a\int_{2\xi}^{\pi}d\varphi_0
E_{\alpha}^{+}(\varphi_0)\\ab\int_{\pi}^{\pi+2\xi}d\varphi_0
E_{\alpha}^{+}(\varphi_0)\\a^{2}b\int_{\pi+2\xi}^{2\pi}d\varphi_0
E_{\alpha}^{+}(\varphi_0)\end{array}\right)\cdot\left(\begin{array}{c}\int^{2\xi}d\varphi
E_{\beta}^{-}(\varphi)\\a^{-1}\int^{\pi}d\varphi
E_{\beta}^{-}(\varphi)\\(ab)^{-1}\int^{\pi+2\xi}d\varphi
E_{\beta}^{-}(\varphi)\\(a^{2}b)^{-1}\int^{2\pi}d\varphi
E_{\beta}^{-}(\varphi)\end{array}\right)
-\left(\begin{array}{c}\int_{0}^{2\xi}d\varphi_0
E_{\alpha}^{+}(\varphi_0)\\\int_{2\xi}^{\pi}d\varphi_0
E_{\alpha}^{+}(\varphi_0)\\\int_{\pi}^{\pi+2\xi}d\varphi_0
E_{\alpha}^{+}(\varphi_0)\\\int_{\pi+2\xi}^{2\pi}d\varphi_0
E_{\alpha}^{+}(\varphi_0)\end{array}\right)\cdot\left(\begin{array}{c}\int^{\varphi_0}d\varphi
E_{\beta}^{-}(\varphi)\\\int^{\varphi_0}d\varphi
E_{\beta}^{-}(\varphi)\\\int^{\varphi_0}d\varphi
E_{\beta}^{-}(\varphi)\\\int^{\varphi_0}d\varphi
E_{\beta}^{-}(\varphi)\end{array}\right)\nonumber\\&=2(\lambda^{1}_{\alpha+}\lambda^{1}_{\beta i}+\lambda^{2}_{\alpha+}\lambda^{2}_{\beta i})-\int_{0}^{2\pi}d\varphi_0
E_{\alpha}^{+}(\varphi_0)\int^{\varphi_0}d\varphi
E_{\beta}^{-}(\varphi)
\end{align}
Integration of the second term then leads to
\begin{eqnarray}
\lambda_{\alpha}^{+}\cdot\lambda_{\beta}^{\textrm{init}}=2(\lambda^{1}_{\alpha+}\lambda^{1}_{\beta i}+\lambda^{2}_{\alpha+}\lambda^{2}_{\beta i})+{\pi\gamma\over1+\gamma^2}\left(
\begin{array}{cc}
 1 & \gamma     \\
 -\gamma &  1\end{array}
\right)_{\alpha\beta}
\end{eqnarray}
and hence
\begin{eqnarray}
\sigma_{\alpha\beta}&=&{\sigma_0\over1+\gamma^2}\left(
\begin{array}{cc}
 1 & \gamma \\ -\gamma & 1\end{array}
 \right)_{\alpha\beta}
+{2\sigma_0\over\pi\gamma}
\left[(\lambda^{1}_{\alpha+}\lambda^{1}_{\beta
     i}+\lambda^{2}_{\alpha+}\lambda^{2}_{\beta
     i})-(\lambda^{1}_{\alpha+}\lambda^{1}_{\beta-}+\lambda^{2}_{\alpha+}\lambda^{2}_{\beta-})
  \right]\nonumber\\ 
&+&{2\sigma_0\over\pi\gamma}\left[
\frac{\lambda^{1}_{\alpha+}\lambda^{1}_{\beta-}(1-bq)+\lambda^{2}_{\alpha+}\lambda^{2}_{\beta-}(1+aq)+\lambda^{1}_{\alpha+}\lambda^{2}_{\beta-}ap-\lambda^{2}_{\alpha+}\lambda^{1}_{\beta-}bp}{1-bq+a(q+b(p^2-q^2))}\right].
\end{eqnarray}
The final result in Eq.~(\ref{conduc}) is obtained by substituting
expressions for
$\lambda_{\alpha\pm}^{1}$ and $\lambda_{\alpha\pm}^{2}$ 
from Eq.~(\ref{lambdaalphapm}) and for
$\lambda_{\alpha i}^{1}$ and $\lambda_{\alpha i}^{2}$
from Eq.~(\ref{lambdaalphainit})  
and
simplifying.
\end{widetext}

\end{document}